\documentclass[preprint2]{aastex}

\slugcomment{To appear in ApJ}

\shorttitle{Modeling Bell's Non-resonant Cosmic Ray Instability}
\shortauthors{V.N. Zirakashvili, V.S.Ptuskin and H.J.V\"olk}

\begin{document}


\title{\bf Modeling Bell's
Non-resonant Cosmic Ray Instability}


\author{V.N.Zirakashvili}
\affil{Pushkov Institute for Terrestrial Magnetism, Ionosphere and Radiowave
Propagation, 142190, Troitsk, Moscow Region, Russia}
\affil{Max-Planck-Institut f\"{u}r\ Kernphysik, Postfach 103980, 69029
Heidelberg, Germany}
\author{V.S.Ptuskin}
\affil{Pushkov Institute for Terrestrial Magnetism, Ionosphere and Radiowave
Propagation, 142190, Troitsk, Moscow Region, Russia}
\and
\author{H.J.V\"olk}
\affil{Max-Planck-Institut f\"{u}r\ Kernphysik, Postfach 103980, 69029
Heidelberg, Germany}



\begin{abstract}
We have studied
the non-resonant streaming instability of charged energetic particles moving
through a background plasma, discovered by Bell
\cite{bell04}. We confirm his numerical results regarding a
significant magnetic field amplification in the system. A detailed physical
picture of the instability
development and of the magnetic field  evolution is given.
\end{abstract}





\section{Introduction}

The diffusive shock acceleration process (Krymsky
\cite{krymsky77}; Axford et al. \cite{axford77}; Bell
\cite{bell78}; Blandford and Ostriker \cite{blandford78}) is
considered as the principal mechanism for the production of the
galactic cosmic rays in supernova remnants (SNRs). A great
strength of the random magnetic fields that provide the scattering
of energetic particles both upstream and downstream of a supernova
shock is necessary for efficient acceleration. It was originally
suggested that this may be the result of a gyroresonant streaming
instability that develops due to the presence of a diffusive
streaming of accelerated particles (Bell \cite{bell78}, Blandford
\& Ostriker \cite{blandford78}).  The corresponding
magnetohydrodynamic (MHD) waves have wavelengths of the order of
the gyroradii of the energetic particles. This kind of resonant
cyclotron instability had been suggested earlier to regulate the
propagation of the galactic cosmic rays (Lerche \cite{lerche67},
Wentzel \cite{wentzel74}). If the generated random fields are
amplified up to wave amplitudes that correspond to the strength
of the mean magnetic field, the scattering mean free path of
resonant energetic particles will decrease to a value comparable
with their gyroradius. This regime of diffusion is called Bohm
diffusion and is often used to estimate the maximum energy of the
accelerated particles.

However, even this rather optimistic regime of diffusion may ensure the
 acceleration of cosmic ray protons only up to energies of about $10^{14}$ eV
 (Lagage \& Cesarsky \cite{lagage83}, Berezhko \cite{ber96}) if the magnetic
 field in the remnant is comparable with the interstellar magnetic field, which
 is typically less than $10$ $\mu $G. Significant magnetic field amplification
 is necessary for acceleration up to higher energies. Amplification might be
 possible since quasilinear theory of the resonant streaming instability allows
 it (see e.g. McKenzie \& V\"olk \cite{mckenzie82}). Since this
 perturbation theory breaks down in the case of resonant interaction of
 particles with high amplitude MHD waves, Lucek and Bell \cite{lucek00}
 performed MHD simulations combined with calculations of energetic particle
 trajectories and found that the magnetic field can be amplified to a level
 where the random field exceeds the initial mean field. The necessity to
 perform detailed trajectory calculations leads to strong numerical limitations
 of such simulations.

Different qualitative treatments of magnetic field amplification
at supernova shocks were suggested on an analytical level (see
e.g. Bell \& Lucek \cite{bell01}, Ptuskin \& Zirakashvili
\cite{ptuskin03}, Vladimirov et al. \cite{vladimirov06}, Amato \&
Blasi \cite{amato06}). Using the observed synchrotron emission,
field amplification was included on a phenomenological level in
numerical solutions of the coupled gas dynamic and particle
acceleration equations (e.g. Berezhko et al. \cite{berezhko02},
Berezhko \& V\"olk \cite{berezhko04a, berezhko04b}, V\"olk et al.
\cite{voelk07}). It became clear that such amplification can lead
to particle accelaration to knee energies at $\approx 3\cdot
10^{15}$ eV and  -- according to speculative extrapolations --
possibly even beyond.

 On the basis of the dispersion relation for collisionless MHD waves derived by
 Achterberg \cite{achterberg83}, Bell \cite{bell04} found a non-resonant
 streaming instability that had been overlooked before. He
 argued that the diffusive streaming of particles accelerated at supernova
 shocks may be so strong that it modifies the dispersion relation of MHD waves
  in an essential way. Then a non-oscillatory purely
 growing MHD mode appears at scales smaller than the gyroradius of the
 particles which excite the instability.  The growth rate of this mode can be
 larger than the growth rate of the resonant mode. Since the particle
 trajectories are only weakly deflected by small-scale magnetic
 inhomogeneities, one can avoid complicated trajectory
 calculations. Only the mean cosmic-ray flux needs to be specified. Bell \cite
 {bell04} performed corresponding MHD simulations and showed that the magnetic
 field can indeed be significantly amplified.

In the following we investigate this non-resonant instability in considerably
more detail. We have performed MHD simulations of the instability with very
good  numerical resolution and have found a simplified
analytical description for the magnetic field amplification. Our results may be
applied to different astrophysical objects, where energetic particles exist.

The present work deals with the instability in the presence of a
  non-resonant energetic particle population streaming through the thermal
  plasma. And it simulates the nonlinear instability development.  In
   a companion paper (Zirakashvili \& Ptuskin \cite{zirakashvili08},
  Paper II) this modeling will be combined with
   an analytical treatment of the diffusive acceleration of
  particles in a plane, steady shock with its
  intrinsically non-uniform energetic particle distribution in the precursor.

The paper is organized as follows: The basic equations are given
in the next section. The scattering of cosmic ray particles by
small-scale random magnetic fields is considered in Sect.3.
The cosmic ray electric current is calculated in Sect. 4.  The
non-resonant instability and the MHD simulations are described in
Sect.5 and 6. The summary is given in the last Section.

\section{Basic equations}

We consider a system that consists of a thermal plasma and a cosmic ray gas
with a negligible mass density.  We shall treat the gas
as a magnetized fluid with frozen-in magnetic field ${\bf B}$ and induced
electric field ${\bf E=-[u\times B]}/c$. Here ${\bf u}$ is the mass velocity
(essentially equal to the thermal gas velocity) and $c$ is the velocity of
light. The electric charge density $\rho _\mathrm{th}$ and electric current
density ${\bf j}_\mathrm{th}$ of the thermal plasma that determine the Lorentz
force ${\bf F}=[({\bf j}_\mathrm{th}-\rho _\mathrm{th}{\bf u})\times {\bf
B}]/c$ may be found from the quasi-neutrality condition $\rho
_\mathrm{th}=-\rho _\mathrm{cr}$ and Amp\`ere's law $[\nabla \times {\bf
B}]=4\pi ({\bf j}_\mathrm{th}+{\bf j}_\mathrm{cr})/c$, respectively. Here $\rho
_\mathrm{cr}$ and ${\bf j}_\mathrm{cr}$ are the electric charge density and
electric current density of the cosmic ray gas, respectively. The Euler
equation of the gas motion may then be written as

\[
\rho \left( \frac {\partial {\bf u}}{\partial t}+({\bf u\nabla )u}\right) =
-\frac {[\bf B\times [\nabla \times B]]}{4\pi }-\nabla P
\]
\begin{equation}
-\frac 1c[({\bf j}_\mathrm{cr}-\rho _\mathrm{cr}{\bf u})\times {\bf B}],
\end{equation}
where $P$ is the thermal gas pressure.

The evolution of the mass density $\rho $ and the magnetic field ${\bf B}$ are
governed by the continuity equation
\begin{equation}
\frac {\partial \rho }{\partial t}+\nabla (\rho {\bf u})=0
\end{equation}
and Faraday's law

\begin{equation}
\frac {\partial {\bf B} }{\partial t}=[\nabla \times [{\bf u}\times {\bf B}].
\end{equation}
The equation for the plasma energy density
$e=(\rho {u^2}/2)+(B^2/(8\pi ))+P/(\gamma -1)$
 may be written as:
\[
\frac {\partial e }{\partial t}+\nabla \left( \rho {\bf u} \frac {u^2}{2}
+\frac {\gamma P {\bf u}}{\gamma -1} +\frac {[{\bf B\times [u\times B]}}{4\pi
}\right)
\]
\begin{equation}
=-\frac {\bf u}{c}[{\bf j}_\mathrm{cr}\times {\bf B}].
\end{equation}
Here $\gamma $ is the adiabatic index of the gas.  The last term in this
equation describes the mechanical work produced by cosmic rays.  The electric
charge density $\rho _\mathrm{cr}$ and the electric current density of cosmic
rays ${\bf j}_\mathrm{cr}$ may be found from the momentum distribution of
cosmic rays $f({\bf p,r},t)$.  It obeys the equation
\begin{equation}
\frac {\partial f }{\partial t}+{\bf v}\nabla f+\frac qc
[({\bf v}-{\bf u})\times {\bf B}]\frac {\partial f }{\partial {\bf p}}=0.
\end{equation}
Here $q$ is the charge of cosmic ray particles.

The system of Eqs. (1)-(5) describes the interaction of cosmic rays and
magnetized thermal plasma. We shall use this system in the next sections.

\section{Scattering by the small-scale field}

The cosmic ray momentum distribution may be written as $f=f_0+\delta f$. Here
$f_0=\left< f\right> $ is the momentum distribution averaged over the
fluctuations of the magnetic field $\delta {\bf B}$ and of the plasma velocity
$\delta {\bf u}$; $\delta f$ is the fluctuation of the momentum
distribution. One can use perturbation theory for the calculation of $\delta f$
when the fluctuations of the magnetic field are small in comparison with the
mean field ${\bf B}_0$. Since we are interested in the investigation of
considerable magnetic field amplification, the theory of perturbations with
small magnetic field amplitude is, generally speaking, not
applicable. Fortunately in the case of the non-resonant instability that is
most interesting for the present consideration, only small-scale fields are
generated (see also below) and the scale of the random field is smaller than
the particle gyroradius in the total magnetic field. Perturbation theory is
applicable in this case because the particles are only weakly deflected on the
characteristic scale of the random field.  The theory of cosmic ray diffusion
in such magnetic fields was developed by Dolginov and Toptygin
\cite{dolginov67}.

The equation for the fluctuation of the momentum distribution can be found from
Eq. (5):

\[
\frac {\partial \delta f }{\partial t}+{\bf v}\nabla \delta f+\frac qc [({\bf
v}-{\bf u}_0)\times {\bf B}_0]\frac {\partial \delta f }{\partial {\bf p}}
\]
\begin{equation}
=-\frac qc [({\bf v}-{\bf u}_0)\times {\bf \delta B}]\frac {\partial f_0
}{\partial {\bf p}}.
\end{equation}
Here ${\bf u}_0$ is the mean  mass velocity and we neglect
the velocity perturbations $\delta {\bf u}$. It is assumed that they
 are small in comparison with
the mean velocity ${\bf u}_0$.
Such perturbations
would only result in
second order Fermi acceleration which is
a factor of $\left< \delta u^2\right> /c^2$ slower
in comparison with particle scattering.

 Since the random field ${\bf \delta B}$ has small spatial scales, the last
term on the left-hand side of this equation may be
also neglected. This means that to lowest order the particles
can be considered to move in straight-line orbits. The
calculations are
in this case significantly simplified in comparison with the
general case, when the integration along the helical orbits of
particles results in the appearance of series containing
Bessel functions (see e.g.  Berezinskii et al.
\cite{berezinsky90}).  Although the zero-order orbits are
different for these two methods, they give similar results in the
case of almost any small-scale random magnetic field. The formal
limit $B_0\to 0$ is not trivial and was considered by Tsytovich
\cite{tsytovich77}.

As long as the gradient lengths of $f_0$ and ${\bf u}_0$ are large compared to
  the spatial fluctuation scale $k^{-1}$, the fluctuation amplitudes
  carry a corresponding parametrical spatial dependence. We shall assume this
  in the following. Fourier transforming then in time and space we find the
  expression for the Fourier transform of the distribution $\delta f_{\omega
  ,{\bf k}}=\int d^3rdt \delta f\exp (i(\omega t-{\bf kr}))(2\pi )^{-4}$:

\begin{equation}
\delta f_{\omega ,{\bf k}}=\frac {q/c}{i(\omega -{\bf kv})} [({\bf v}-{\bf
u}_0)\times {\bf \delta B_{\omega ,k}}]\frac {\partial f_0 }{\partial {\bf p}}.
\end{equation}

Here ${\bf \delta B_{\omega ,k}}$ is the Fourier transform of the random
magnetic field. We shall  further assume that the
magnetic field changes slowly in time in the frame moving with the mean
mass velocity ${\bf u}_0$ and therefore write ${\bf \delta B_{\omega
,k}}=\delta {\bf B_k}\delta (\omega -{\bf ku}_0)$.

Averaging now Eq. (5) and using Eq.(7)  we
obtain the following kinetic equation for the average cosmic ray
distribution function $f_0$:

\[
\frac {\partial f_0 }{\partial t}+{\bf v}\nabla f_0+q\left( {\bf E}_0+\frac 1c
[({\bf v}-{\bf u}_0)\times {\bf B}_0]\right) \frac {\partial f_0 }{\partial
{\bf p}}
\]

\begin{equation}
=\frac \partial {\partial p_i}\nu _{ij}\frac {\partial f_0}{\partial p_j}
\end{equation}
Here ${\bf E}_0=-c^{-1}\left< \bf \delta u\times \delta B\right> $ is the mean
electric field in the frame of reference moving with the mean plasma
velocity ${\bf u}_0$.

 The scattering tensor $\nu _{ij}$ in the last equation  is determined by the
spectrum of the random magnetic field $B_{ij}({\bf k})=\int \frac {d^3r}{(2\pi
)^3}\left< B_i{\bf (r+r}_0)B_j{\bf (r}_0)\right> \exp (-i{\bf kr)}$:

\[
\nu _{ij}=\frac {q^2}{c^2}\pi \int d^3k\delta ({\bf
k(v-u}_0))e_{ilm}e_{jrs}
\]
\begin{equation}
\times B_{lr}({\bf k})
(v_m-u_{0m})(v_s-u_{0s})
\end{equation}
Here $e_{ijk}$ is the antisymmetric tensor. The scattering tensor makes the
cosmic ray distribution isotropic in the frame moving with the velocity ${\bf
u}_0$. Expression (9) may be simplified in the case
of an isotropic random magnetic field $B_{ij}({\bf k})=\frac
12B_{\mathrm{isotr}}(k)(\delta _{ij}-k_ik_j/k^2)$ (Dolginov \& Toptygin
\cite{dolginov67}):
\begin{equation}
\nu _{ij}=p^2\nu (p) \frac
{({\bf v-u_0})^2\delta _{ij}-(v_i-u_{0i})(v_j-u_{0j})}
{2v|{\bf v-u}_0|}
\end{equation}
where the scattering frequency $\nu (p)$ is given by the formula
\begin{equation}
\nu (p)=\frac \pi 4\frac {q^2v}{p^2c^2}\int d^3k B_{\mathrm{isotr}}(k)/k
\end{equation}
Here the spectrum of the isotropic magnetic field is normalized as $\left< \delta
B^2\right> =\int d^3kB_{\mathrm{isotr}}(k)$.

In the diffusion approximation the  average cosmic ray
distribution function $f_0$ may be written as

\begin{equation}
f_0({\bf p})=N(p)+\frac 3{pv}{\bf pJ}(p),
\end{equation}
where $N(p)$ is the isotropic part and ${\bf J}(p)$ is the cosmic ray flux density,
which is the sum of the diffusive and advective flux densities and is given by
Eq. (A.3) in Appendix A.

\section{Calculation of the electric current}

We shall use Eq. (7) for the calculation of the electric current density of
  the cosmic ray gas.  Substitution of the expression
 (12) into Eq. (7), multiplication by $q{\bf v}$, and integration over momentum
 space give the fluctuating part of the Fourier transform of the cosmic ray
 electric current $\delta {\bf j}_\mathrm{cr, \omega ,{\bf k}}$. Disregarding
 terms of the order $u_0/v$ we obtain

\begin{equation}
\delta {\bf j}_\mathrm{cr, \omega ,{\bf k}}=
3\pi \frac {q^2}c\delta (\omega -{\bf ku}_0)\int \frac {d^3p}{pv}\delta ({\bf
kv)(J}_\mathrm{d}[{\bf v\times \delta B_k])v}.
\end{equation}
Here ${\bf J}_\mathrm{d}={\bf J+u}_0\frac p3\frac {\partial
N}{\partial p}$ is the average cosmic ray diffusion flux. The
appearance of the $\delta $-function in this equation is due
to the Landau resonance $\omega ={\bf kv}$ (cf. Lifshitz and
Pitaevskii \cite{lifshitz81}) in Eq.(7).

 The total electric current of the cosmic  ray gas is
the  flux {$\bf J$} multiplied by the particle charge $q$.

Performing the integration on the two angles in momentum space,
and
  calculating the inverse Fourier transformation  of the last
  equation, we obtain the expression for the diffusive electric current ${\bf
  j}_\mathrm{d}={\bf j}_\mathrm{cr}-\rho _\mathrm{cr}{\bf u}$ that appears in
  the right-hand side of Eq. (1):

\[
{\bf j}_\mathrm{d}=q\int d^3p{\bf J}_\mathrm{d}+
\frac {3\pi}4 \frac {q^2}c\int d^3k\exp (i{\bf k(r-u}_0t))
\]
\begin{equation}
\times \int \frac {d^3p}{pk^3} \left( k^2{\bf [\delta B_k\times J}_\mathrm{d}]-{\bf
k(k[\delta B_k\times J}_\mathrm{d}])\right) .
\end{equation}

The first term on the right-hand-side of this equation is simply
the zero-order term of the expansion in the magnetic fluctuation
$\delta {\bf B}$, while the second term is the linear term of the
expansion. The latter is always smaller than the first term if the
small-scale field approximation is valid, that is if $qB/pck<<1$,
which means that the particle gyroradius $pc/qB$ is large compared
to the scale $k^{-1}$ of the magnetic field.


However, in some cases the second integral may play a r\^ole. When the cosmic
ray streaming is not strong, it results only in a small change of the
dispersion relation of MHD waves. The first integral in Eq. (14) in this case
produces only a small shift of the frequency of MHD waves. The second term then
gives a small imaginary part of the frequency and describes a resonant wave
instability based on the Landau resonance in Eq. (7). Within the limits of the
small-scale field approximation,  and for strong cosmic ray streaming,
this resonant instability is ineffective in comparison to the well-known
gyroresonant streaming instability.

Another important point is that the second integral on the right-hand side of
Eq. (14) appears in the calculation of the mean force acting on the thermal
plasma (Ptuskin \cite{ptuskin84}):
${\bf F}=c^{-1}\left< [{\bf B\times (\bf j}_\mathrm{cr}-\rho _\mathrm {cr}{\bf
u}_0)]\right> $.

Using Eq. (14) and  averaging we obtain

\[
{\bf F}=\frac \pi 4\frac {q^2}{c^2}\int \frac {d^3p}{p}{\bf J}_d\int d^3k B_{\mathrm{isotr}}(k)/k
\]
\begin{equation}
=\int {d^3p}\frac {p}v\nu (p){\bf J}_d
\end{equation}
It was assumed here that the random field is isotropic and
expression (11) was used.  Since the diffusive flux is equal to
${\bf J}_d=-\frac {v^2}{3\nu }\nabla N$ in the diffusion
approximation, ${\bf F}=-\nabla P_\mathrm{cr}$, where
$P_\mathrm{cr}=\int d^3p pvN(p)/3$ is the cosmic ray pressure. In
the sense of MHD theory, Eq. (1) then describes the overall
momentum balance of the system, where the forces on the r.h.s. are
the Lorentz force and the gradient of the overall pressure,
thermal plus nonthermal. Indeed $c^{-1}\left< [({\bf
j}_\mathrm{cr}-\rho _\mathrm{cr}{\bf u}_0) \times {\bf B}]\right>
= -\nabla P_\mathrm{cr}$ is the general form of the average
cosmic ray momentum balance.

In the sequel we will neglect the 2nd term of Eq. (14) (see also Bell
\cite{bell04}). This is justified in the linear analysis and in the
nonlinear simulation of the non-resonant instability if the small-scale field
approximation is valid.

\begin{figure}[t]
\includegraphics[width=7.5cm]{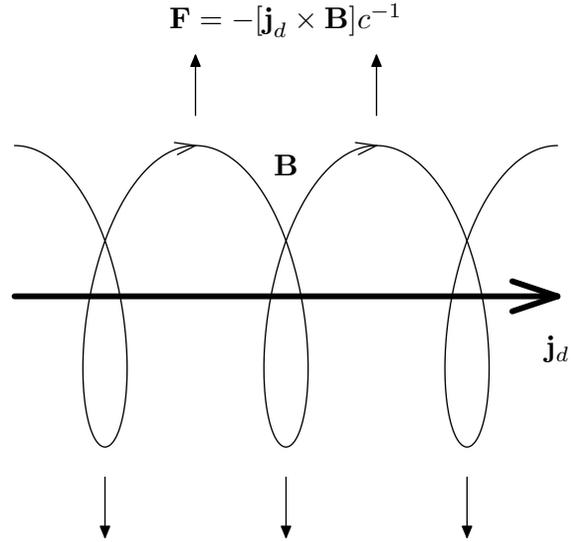}
\caption{Explanation of the non-resonant streaming instability. The magnetic
spiral (thin solid line) is stretched by the Lorentz force ${\bf F}=-[{\bf
j}_\mathrm{d}\times {\bf B}]c^{-1}$ that appears due to
the diffusive cosmic ray electric current ${\bf j}_\mathrm{d}$.}
\end{figure}

\section{Non-resonant streaming instability}

The dispersion relation for small-scale MHD perturbations may be
found from Eqs. (1)-(5). In the case when the
cosmic ray diffusion flux ${\bf J}_\mathrm{d}$ and the wavenumber
${\bf k}$ are parallel to the mean magnetic field ${\bf B}_0$ this
dispersion relation may be written as (Bell \cite{bell04}):
\begin{equation}
(\omega -{\bf ku}_0)^2=V^2_\mathrm{a}k^2\mp \frac {j_\mathrm{d}B_0}{c\rho _0}k.
\end{equation}
Here $\rho _0$ is the mean plasma density, $V_\mathrm{a}=B_0/\sqrt{4\pi \rho
_0}$ is the Alfv\'en velocity,
${\bf j}_\mathrm{d}=q\int d^3p{\bf J}_\mathrm{d}$ is the average
diffusive electric current of cosmic rays,
and the two signs $\mp $ correspond to the two
circular polarizations. A non-resonantly unstable MHD mode appears if the
condition $k< k_\mathrm{c}=j_\mathrm{d}B_0/c\rho _0V^2_\mathrm{a}$ is
fulfilled. The unstable magnetic field line spiral expands in the direction
perpendicular to the mean magnetic field (see Fig.1). The mode with
$k=k_\mathrm{c}/2$ has the maximum growth rate $\gamma _{\max }$ :

\begin{equation}
\gamma _{\max }=\frac {j_\mathrm{d}B_0}{2c\rho _0V_\mathrm{a}},
\end{equation}
which does not depend on the magnetic field strength. 

Since we assumed that the scale $k^{-1}$ of the perturbations is smaller than
the gyroradius {\bf $pc/qB_0$} of the energetic particles,
the wavenumber $k$ should obey the condition $qB_0/pc<< k_\mathrm{c}$. This
means that the necessary condition for instability is
$k_\mathrm{c}> qB_0/pc$. This condition may be rewritten as
\begin{equation}
\frac {u_\mathrm{cr}}{v}\frac {\epsilon _\mathrm{cr}}{B^2_0/4\pi }>> 1
\end{equation}
Here $u_\mathrm{cr}$ and $\epsilon _\mathrm{cr}$ are the bulk velocity and the
energy density of the cosmic ray gas, respectively, and $v$ is the velocity of
energetic particles.  This condition is easily fulfilled at the shocks of SNRs,
where $u_\mathrm{cr}$ is of the order of the shock velocity $u_1$ and the
energy density of the relativistic particles may be comparable with $\rho
u^2_1$.

\begin{figure}
\includegraphics[width=7.5cm]{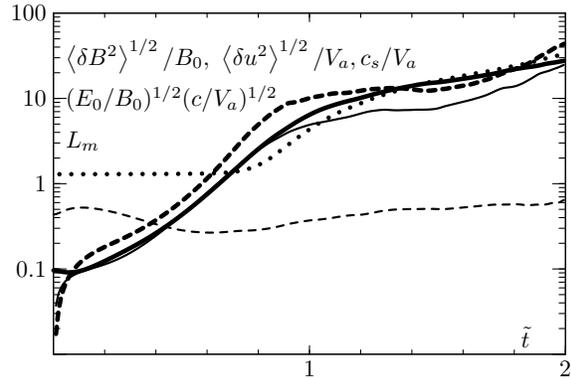}
\caption{ Numerical results of the modeling of the non-resonant instability
with the dimensionless cosmic ray electric current $J=16$ as a function of
normalized time $\tilde{t}$. The r.m.s.  values of the magnetic field
fluctuation $\left< \delta B^2\right> ^{1/2}$ divided by the mean magnetic
field strength $B_0$ and of the mass velocity fluctuations $\left< \delta
u^2\right> ^{1/2}$ divided by the Alfv\'en velocity $V_a$ are shown by the {\it
solid} and {\it dashed lines}, respectively. The ratio of the sonic velocity
$c_\mathrm{s}$ and the Alfv\'en velocity $V_a$ is shown by the {\it dotted
line}. The square root of the ratio of the mean electric field $E_0$ and the
mean magnetic field $B_0$ multiplied on $(c/V_a)^{1/2}$ is shown by the {\it
thin solid line}. The characteristic scale $L_m$ of the magnetic field is also
shown by the {\it thin dashed line}.
}

\end{figure}

\begin{figure}
\includegraphics[width=7.6cm]{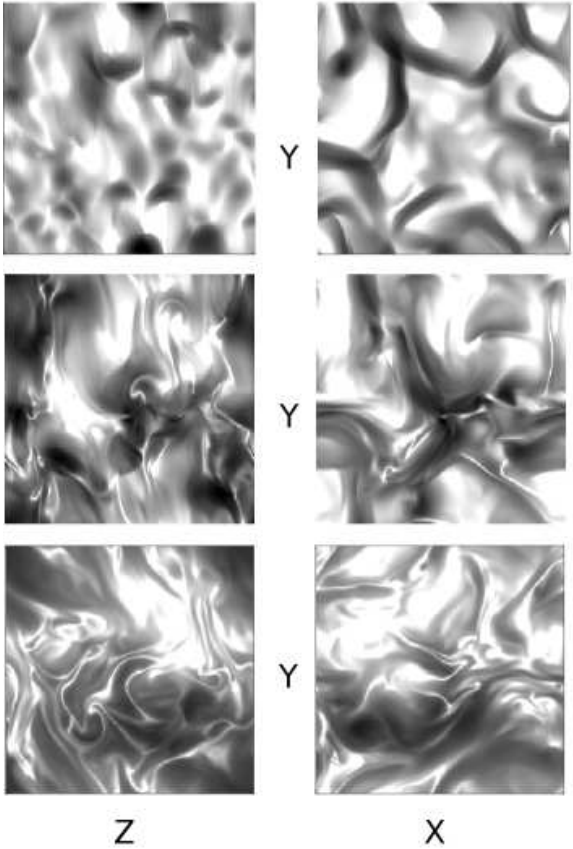}
\caption{Slices of the magnetic field strength through the center of the box,
obtained at $\tilde{t}=0.82$ (top), $\tilde{t}=1.03$ (middle) and
$\tilde{t}=1.48$ (bottom), in the run with $J=16$. The scaling of the magnetic
field strength is logarithmic between $0.32\left< B^2\right> ^{1/2}$ (white) and
$3.2\left< B^2\right> ^{1/2}$ (black). The mean
magnetic field and the diffusive electric current are in the $z$ direction.}
\end{figure}

\section{Numerical modeling of the non-resonant instability}

We  have numerically modeled
 the non-resonant instability similar in spirit to the  modeling of
Bell \cite{bell04}. The MHD Eqs. (1)-(4), written
in dimensionless form, were solved numerically. We used the numerical method of
Pen et al.  \cite{pen03}. It is a  second order in
space and time, flux-conservative total variation diminishing
MHD scheme which enforces the $\nabla {\bf B}=0$ constraint to
machine precision. The nonlinear flux limiter "minmod" was used.

\begin{figure}[t]
\includegraphics[width=8.0cm]{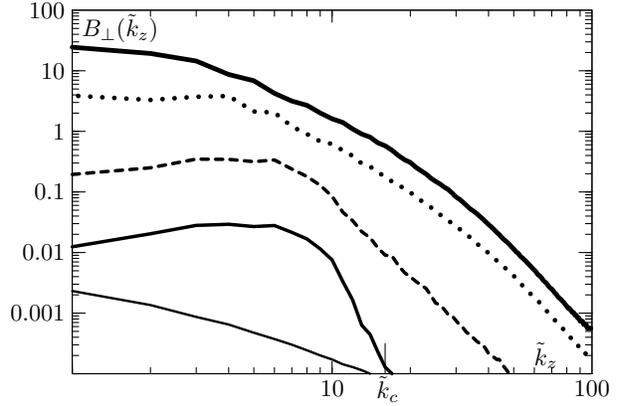}
\caption{The one-dimensional spectra of the perpendicular component of
the random magnetic field $B_\perp (\tilde{k}_z)$ obtained at $\tilde{t}=0$ (thin solid
line),  $\tilde{t}=0.54$ (solid
line), $\tilde{t}=0.74$ (dashed line), $\tilde{t}=0.97$ (dotted line) and
$\tilde{t}=1.26$ (thick solid line).
All spectra are normalized to the square of the mean magnetic field. The
critical wavenumber $\tilde{k}_c=16$ is marked on the $x$-axis.}
\end{figure}

The dimensionless time $\tilde{t}$, the space coordinate $\tilde{z}$ and the
velocity $\tilde{u}$ are  defined as
$\tilde{t}=tV_\mathrm{a}k_0$, $\tilde{z}=k_0z$,
$\tilde{u}=u/V_\mathrm{a}$, respectively. Here $k_0$ is the wavenumber that corresponds to
the real size of the  numerical box $2\pi /k_0$. The dimensionless density
$\tilde{\rho }$ and the electric current $J$ can be expressed via the magnetic
field $B_0$ and the Alfv\'en velocity $V_\mathrm{a}$ as $\tilde{\rho }=4\pi \rho
V_\mathrm{a}^2/B_0^2$ and $J=4\pi j/ck_0B_0$. The dimensionless wavenumber
$\tilde{k}$ is simply $\tilde{k}=k/k_0$.

The simulations were performed in a cubic box with size
$2\pi $. Periodic boundary conditions were imposed at the sides of
the box. We used $256^3$ grid cells in our simulations (to be
compared with $128^3$ cells in Bell's case).

At $\tilde{t}=0$ the plasma pressure and density are uniformly distributed in
space. Small random magnetic perturbations corresponding to isotropically
distributed Alfv\'en waves with a one dimensional spectrum $\propto k^{-1}$ and
$\left<  B^2\right> ^{1/2}/B_0=0.09$
were added to the mean unit
strength magnetic field that is in $z$ direction. Here $\left<  \right>  $
denote the spatial average over the simulation volume.
We use the values
$\gamma =5/3$ and $\beta =1$. Here $\beta = 4\pi P/B_0^2$.

The evolution of the magnetic field and the mass velocity fluctuations,
together with the evolution of the sound speed, the mean electric field $E_0$
and  the characteristic scale $L_m$ of the magnetic field, for a
dimensionless cosmic ray current $J=16$, is shown in Fig.2. The characteristic
scale $L_m=\int dk_zB_{\perp }(k_z)|k_z|^{-1}/\int dk_zB_{\perp }(k_z)$ is
determined via the spectrum $B_{\perp }(k_z)$ of the perpendicular component of
the random magnetic field $B_{\perp }(k_z)=\int dk_xdk_y(B_{xx}({\bf
k})+B_{yy}({\bf k}))$.

After a brief initial stage the fluctuations grow exponentially with a growth
rate that is slightly smaller than $\tilde{\gamma }_{\max }=8$ in dimensionless
units. The initial growth of the magnetic fluctuations is not exponential,
since only  a part of the initial perturbation corresponds to unstable
modes.

At $\tilde{t}=0.7$ the magnetic perturbations are already comparable with the
mean magnetic field.  Parts of the magnetic spiral expanding into the
$XY$-plane begin to collide with their surroundings.
When this happens multiple shocks are formed. The shape of these shocks may be
seen in the top panel of Fig.3, where the magnetic field strength in
perpendicular slices through the center of the box are shown.  In the
$YZ$-plane, which contains the initial magnetic field vector, they look like
bow shocks. The shocks corresponding to the adjacent turns of the same
magnetic spiral are clearly seen.  These shocks are almost
circular in the $XY$-plane. Low density cavities appear inside these
shocks. The size of these cavities in the $XY$ plane is larger than the size in
the $z$ direction.  At later times the shocks collide with each other in the
$XY$ plane and the gas motion becomes strongly turbulent (middle and bottom
panels of Fig.3).

The MHD turbulence has rather small spatial scales in the early stage of the
magnetic field growth. The scale of the magnetic field increases with time (see
Fig.2). This increase is slower in comparison with what was suggested by Bell
$k\sim 4\pi j_\mathrm{d}/Bc$ (or $\tilde{k}\sim JB_0/B$ in dimensionless units)
when the magnetic  tension forces are comparable with the
Lorentz force produced by the cosmic ray electric current. This effect is
illustrated in Fig.4, where the magnetic spectra obtained for several instants
of time, are shown. It is clear from this figure that a nonlinear transfer of
magnetic energy takes place in the system. The increase of the magnetic energy
at wavenumbers $\tilde{k}>\tilde{k}_c$ that are not excited in the linear
approximation demonstrates the non-linear transfer of energy to smaller scales.
The magnetic energy in the small wavenumbers also grows faster than predicted
by the analytical growth rate formula. For example the amplitude of the
harmonics with wavenumber $\tilde{k}=1$ increases by a factor of 4.4 during the
period from $\tilde{t}=0.74$ up to $\tilde{t}=0.97$ (see Fig.4).  The linear
growth rate for this harmonics $\tilde{\gamma
}=\sqrt{J\tilde{k}-\tilde{k}^2}=\sqrt{15}$ corresponds to an amplification
factor of about 2.2 during this period. This demonstrates the nonlinear
transfer of energy to larger scales.  Thus the magnetic energy is non-linearly
transferred to  both smaller and larger scales compared to the linearly excited
spatial scales.

Towards the end of the simulation, at $\tilde{t}\sim 1.5$, the scale of the
magnetic field is comparable with the size of the box. At this point in time
the internal energy of the thermal gas is roughly equal to the kinetic and
magnetic energy in our simulation. Bell continued his calculation beyond this
point and found that at later times the magnetic field reaches a saturation
value.  Also continuing our simulations we found that the magnetic field
continues to grow. We believe that this difference is due to the fact that
different MHD codes were used. The difference is not important however, since
the simulation does not model the real situation any more when the scale of the
field has become comparable with the size of the spatial simulation domain. In
this sense our results are similar to Bell's results.

The growth of the MHD perturbations decreases with increasing magnetic field
amplification. As estimated in Appendix B, the magnetic field is amplified only
linearly at large times, $B/B_0\sim J\tilde{t}$ in dimensionless units, whereas
the gas thermal energy density increases $\propto t^3$. These dependencies are
derived using the equation for the evolution of the magnetic helicity.

\section{Conclusion}

We have  modeled the non-resonant
instability produced by a flux of charged
energetic particles that is driven through a scattering thermal
plasma. Using a significantly better numerical resolution we
 basically confirm the results obtained earlier by Bell
\cite{bell04}. The magnetic field may be amplified significantly. The unstable
magnetic spirals collide with each other and shocks of moderate strength are
formed  as the instability
develops (see previous Section). These shocks lead to significant gas heating.
Since free expansion of the magnetic
spirals after collision is impossible, the field grows only linearly in time at
later epochs (see Eq. (B3)). If the system has enough time to
evolve, the magnetic field growth will be stopped when the gyroradius $pc/qB$
of the energetic particles in the amplified field $B$ will drop down to the
scale of the amplified field $k^{-1}\sim cB/4\pi j_\mathrm{d}$. This determines
the value of the saturated magnetic field (Bell
\cite{bell04}, Pelletier et al. \cite{pelletier06}):

\begin{equation}
\frac {B^2}{4\pi }\sim \frac {u_\mathrm{cr}}{v}\epsilon _\mathrm{cr}.
\end{equation}

We should note that the small-scale approximation considered in Sect. 3
becomes invalid when the magnetic field reaches
this saturation value.

Since the instability is driven by the Lorentz force, the corresponding MHD
turbulence has specific properties. It has non-zero magnetic helicity and
a non-zero mean electric field ${\bf E}_0$ parallel to the mean magnetic
field.

We expect that the formation of multiple shocks in three-dimensional
MHD turbulence will also occur for other instabilities, in particular
for the resonant streaming instability, driven by cosmic rays.

The scattering of energetic particles by the small-scale magnetic
inhomogeneities can be described using the Dolginov-Toptygin
approximation (Dolginov \& Toptygin \cite{dolginov67}). The
appearance of a mean second order electric field
${\bf E}_0=-c^{-1}\left< \bf \delta
u\times \delta B\right> $, which is
oppositely directed to the electric current of the energetic
particles, modifies the cosmic ray transport equation (see
Appendix A).

This non-resonant streaming instability may be important in any
astrophysical site where a strong electric current of energetic particles
exists and where the initial magnetic strength is small enough (see condition
(18)). Supernova remnants, starburst galaxies, galaxy cluster accretion shocks
and AGN jets are  possible candidates for an application
of this instability.

The  results  obtained will be used
in a companion paper by Zirakashvili \& Ptuskin \cite{zirakashvili08} (Paper
II) for a  model of diffusive shock acceleration
in young SNRs in the presence of the non-resonant streaming
instability.

\begin{acknowledgements}
We thank the anonymous referee for a number of valuable comments.
VSP and VNZ acknowledge the hospitality of the Max-Planck-Institut f\"ur
Kernphysik, where this work was mainly carried out. The work was also supported
by the RFBR grant in Troitsk.

\end{acknowledgements}

\appendix
\section{Diffusion approximation in the presence of the
   additional electric field}

The cosmic ray transport equation is modified in the presence of
the additional mean electric field ${\bf E}_0=-c^{-1}\left< \bf \delta
u\times \delta B\right> $ (Fedorov et al. \cite{fedorov92}).
Let us substitute the cosmic ray momentum
distribution (12) into Eq. (8) with the scattering tensor (10).
Performing the expansion up to the second order in $u/v$ and
collecting the terms independent of the direction of the particle
velocity ${\bf v}$ and  separately those proportional to the
velocity we obtain after some algebra

\begin{equation}
\frac {\partial N}{\partial t}+\nabla {\bf J}+ \frac {1}{p^2}\frac
{\partial }{\partial p}\frac {p^2}{v}q({\bf E}_0{\bf J}) =\frac
{{\bf u}_0}{p^2}\frac {\partial }{\partial p}\frac {p^3}{v^2}
\left( \nu \left( {\bf J+u}_0\frac {p}{3}\frac {\partial
N}{\partial p}\right) +\left[ {\bf \Omega \times J}\right] \right)
,
\end{equation}

\begin{equation}
\frac {\partial {\bf J}}{\partial t}+\frac {v^2}{3}\nabla N +q{\bf
E}_0\frac {v}{3}\frac {\partial N}{\partial p} = -\nu \left( {\bf
J+u}_0\frac {p}{3}\frac {\partial N}{\partial p}\right) + \left[
{\bf \Omega }\times \left( {\bf J+u}_0\frac {p}{3}\frac {\partial
N}{\partial p}\right) \right]
\end{equation}

Here ${\bf \Omega }=qv{\bf B}_0/pc$. Assuming a slow time evolution we
neglect the time derivative in the last equation. Then it can be
used to find the cosmic ray flux ${\bf J}$:
\begin{equation}
J_i=-D_{ij}\left( \nabla _jN+E_{0j}\frac {q}{v}\frac {\partial
N}{\partial p}\right) -{u_{0i}}\frac {p}{3}\frac {\partial
N}{\partial p},
\end{equation}
where the diffusion tensor $D_{ij}$ has the following form
\begin{equation}
D_{ij}=(D_{\parallel }-D_{\perp })b_ib_j+D_{\perp }\delta
_{ij}+D_Ae_{ijk}b_k.
\end{equation}
Here ${\bf b}={\bf B}_0/B_0$ is the unit vector in the direction of the mean
field ${\bf B}_0$, $D_{\parallel }$ and $D_{\perp }$ denote the parallel and
perpendicular diffusion coefficients, respectively, and $D_A$ is the
antisymmetric diffusion coefficient. They are given
by Dolginov \& Toptygin {\cite{dolginov67}):
\begin{equation}
D_{\parallel }=\frac {v^2}{3\nu }, \ D_{\perp }=\frac {v^2\nu
/3}{\Omega ^2+\nu ^2}, \ D_{A}=\frac {v^2\Omega /3}{\Omega ^2+\nu
^2}.
\end{equation}
Then Eq. (A1) reduces to
\begin{equation}
\frac {\partial N}{\partial t}+{\bf u}_0\nabla N-\frac {p}{3}\frac
{\partial N}{\partial p}\nabla {\bf u}_0 =\left( \nabla _i+\frac
{1}{p^2}\frac {\partial }{\partial p}\frac {p^2q}{v}E_{0i} \right)
D_{ij}\left( \nabla _jN+\frac qvE_{0j}\frac {\partial N}{\partial
p} \right) .
\end{equation}
Eq. (A6) shows how the presence of the mean electric field ${\bf E}_0$ modifies
the diffusion term of the cosmic ray transport equation.

\section{Magnetic helicity}

As noted first by Pelletier et al. \cite{pelletier06}, the MHD
turbulence generated by the non-resonant instability has nonzero
magnetic helicity $H=\left< \delta {\bf A} \delta {\bf B}\right> $,
where $\delta {\bf A}$ is the perturbation of the magnetic
potential. The magnetic helicity $H$ is a useful quantity and it
is often used in the theory of MHD turbulence and dynamo theories
(see e.g. Biskamp \cite{biskamp03}). Faraday's equation (3) may be
used for the determination of the time evolution of this quantity.
For the periodic system we have (cf. Subramanian \&
Brandenburg \cite{subramanian04})
\begin{equation}
\frac {\partial H}{\partial t}=-2c{\bf B}_0{\bf E}_0
\end{equation}

We neglect magnetic dissipation here. This seems well justified because the
magnetic helicity is an integral quantity and is not transferred by nonlinear
interactions to smaller scales where dissipation is essential. In this sense
the magnetic helicity is different from the nonthermal (kinetic + magnetic)
energy that may be transferred to smaller and smaller scales where it is
transformed into gas thermal energy even in the case of infinitely small
viscosity.

The mean electric field appears in the system as a response of the medium to
the cosmic ray electric current. The evolution of the energy density of the
plasma takes place according to Eq. (4).  The right-hand side of this equation
is simply $-{\bf E}_0{\bf j}_\mathrm{d}$. Comparing with Eq. (B1) for a time
independent $j_\mathrm{d}$, we obtain the relation
\begin{equation}
H-2\frac {B_0\left< e\right> }{j_\mathrm{d}}=\mathrm{const}(t)
\end{equation}
It is worth emphasizing that Eq. (B2) is valid for the total plasma energy
density $e$ which appears in Eq. (4).  During the stage of exponential growth
the instability produces magnetic field and velocity perturbations, whereas gas
heating is important at later times.

We may use Eq. (B1) for a derivation of the equation for the magnetic field
amplification. Since the turbulence is helical, the magnetic helicity (that is
the product of the magnetic field and the vector magnetic potential) is $H\sim
B^2/k$, the electric field is $E_0\sim V_\mathrm{a}B^2/cB_0$, and $k\sim 4\pi
j_\mathrm{d}/Bc$. Then Eq. (B.1) yields the equation for the amplification of
the magnetic field:

\begin{equation}
\frac {\partial B}{\partial t}\sim \frac {4\pi V_\mathrm{a}j_\mathrm{d}}{c}
\end{equation}
The numerical factor in this equation is of order unity, according to our
numerical results.

There is another way to obtain this last equation. The field is amplified by
turbulent motions of the medium, that is $\partial B/\partial t\sim B_0ku_k$,
where $u_k$ is the turbulent velocity with wavenumber $k$. Assuming
equipartition, $u_k\sim V_\mathrm{a}B/B_0$, and the estimate $kB\sim 4\pi
j_\mathrm{d}/c$, we arrive at Eq. (B3).

Since $H\propto B^3$ at late times, also $\left< e\right> \propto
t^3$, cf. Eq. (B2). On the other hand $B^2\propto t^2$, and
therefore at very late times the gas internal energy dominates.

\end{document}